# ARTICLE

# Workflow for practical quantum chemical calculations with quantum phase estimation algorithm: electronic ground and π-π* excited states of benzene and its derivatives†

Yusuke Ino,*[a] Misaki Yonekawa,[a] Hideto Yuzawa,[a] Yuichiro Minato,[b] and Kenji Sugisaki [cde]

Quantum computers are expected to perform the full-configuration interaction calculations with less computational resources compared to classical ones, thanks to the use of the quantum phase estimation (QPE) algorithms. However, only a limited number of the QPE-based quantum chemical calculations have been reported even for numerical simulations on a classical computer, and the practical workflow for the QPE computation has not yet been established. In this paper, we report the QPE simulations of the electronic ground and the π–π* excited singlet state of benzene and its chloro- and nitro-derivatives as the representative industrially important systems, with the aid of GPGPU acceleration of quantum circuit simulations. We adopted the pseudo-natural orbitals obtained from the MP2 calculation as the basis for the wave function expansion, the CISD calculation within the active space to find the main electronic configurations to be included in the input wave function of the excited state, and the technique to reduce the truncation error the calculated total energies. The proposed computational workflow is easily applicable to other molecules and can be a standard approach for performing the QPE-based quantum chemical calculations of practical molecules.

## 1. Introduction

In recent years, quantum computing has emerged as a ground-breaking technology that has attracted widespread attention as the next-generation frontier. The growing complexity of modern technological challenges, particularly in fields such as chemistry, materials science, and finance, has surpassed the capabilities of conventional classical computers. Quantum computing is promising for addressing these challenges, which were previously considered almost impossible to solve.[1–9] Consequently, numerous countries and organizations are actively exploring and investing in this transformative technology. Several hardware architectures for quantum bits (qubits) have already been proposed[10] (e.g. superconducting circuits, neutral atoms, trapped ions, and photonic devices). Currently available quantum hardware is, however, noisy and of intermediate scale one, and its application to real-world problems is quite challenging.[11,12]

It should be emphasised that the progress in the development of quantum hardware towards fault-tolerant quantum computing (FTQC) is remarkable. For example, experimental demonstration of 48 logical qubit systems based on reconfigurable neutral atom arrays has been reported recently.[13] Considering such rapid progress in quantum hardware, the research of the FTQC algorithms is an urgent issue.

Among the research fields where quantum computers are expected to bring computational advantages, quantum chemical calculations are considered as one of the most promising applications.[14–17] Quantum chemical calculations are widely used to analyse and predict the molecular structures, electronic states and molecular properties, and reaction mechanisms of molecules. They are applied to various fields such as the design and understanding of light-absorbers,[18] the prediction of reactants and the discovery of new reactions in catalyst development,[19] the analysis of interactions on conducting organic polymers of cathodic half-cells with surrounding species and the resulting charge distribution changes,[20] and the analysis of metabolomics of small biomolecules.[21] As their applications increase, so does their role as a theoretical foundation for the development of new materials. The importance of their potential applications as a tool for obtaining the design principles of new materials prior to synthesis has also been widely recognised. However, accurate and reliable predictions of the molecular properties of unknown materials remain challenging. To date, density-functional theory calculation[22] is widely used to investigate electronic structures of molecules, but it potentially suffers from arbitrariness of the choice of exchange–correlation

[a.] *Fujifilm Corporation, 577 Ushijima, Kaisei-cho, Ashigarakami-gun, Kanagawa 258-8577, Japan. E-mail: yusuke.ino@fujifilm.com*
[b.] *blueqat Inc., 2-24-12-39F, Shibuya, Shibuya-ku, Tokyo 150-6139, Japan.*
[c.] *School of Science and Technology, Keio University, 7-1 Shinkawasaki, Saiwai-ku, Kawasaki, Kanagawa 212-0032, Japan*
[d.] *Quantum Computing Center, Keio University, 3-14-1 Hiyoshi, Kohoku-ku Yokohama, Kanagawa 223-8522, Japan*
[e.] *Centre for Quantum Engineering Research and Education, TCG Centres for Research and Education in Science and Technology, Sector V, Salt Lake, Kolkata 700091, India*

† Electronic Supplementary Information (ESI) available: Cartesian coordinates of LiH, benzene, chlorobenzene, and nitrobenzene, active orbitals used for the CAS-CI and the IQPE calculations, details of the size of problems including number of qubits, Hamiltonian terms, and quantum gates, the quantum circuit used for the excited state approximate wave function preparation. See DOI: 10.1039/x0xx00000x

functionals[23]. It is desirable to use sophisticated ab-initio molecular orbital theories for the calculations of unknown molecules. The most reliable ab initio method for calculating molecular properties is the full-configuration interaction (full-CI) method,[24] which fully considers electronic correlations by including all mixings of electronic configurations. However, the required computational resources for the full-CI calculation increases exponentially with the size of the molecule and it is impractical even for medium-size molecules. To tackle such situations, the full-CI treatments within an active space such as the complete active space self-consistent field (CASSCF) and the CAS-CI are widely used. These approaches are efficient to consider static electronic correlation effects to describe the electronic structures of strongly correlated systems, such as molecules undergoing covalent bond dissociation and multi-nuclear transition metal complexes with antiferromagnetic exchange couplings. [25,26] Again, the computational cost for the CASSCF and the CAS-CI grows exponentially with the size of active spaces, and therefore acceleration of the calculations is highly desirable. In a quantum computer, we can express such mixing of electronic configurations by means of quantum superposition state of the $N^{so}$ quantum bits, where $N^{so}$ is the number of spin orbitals. The mixing of configurations under a given Hamiltonian is simulated by interference of quantum states. The result of the calculation can be retrieved through the measurements of qubits after the quantum circuit operations. This means that quantum computers could mimic the desired electronic states with exponentially fewer resources than classical computers ("exponential acceleration"), under certain conditions.

Currently, most of the reported computational results relevant to quantum chemical calculations on quantum computers are based on the variational quantum eigensolver (VQE) algorithms.[27-29] Although the VQE can be implemented in currently available noisy intermediate-scale quantum (NISQ) computers, recent studies also revealed its challenges to overcome, such as shot noises on the energy expectation values,[30] hardness of the variational optimizations,[31] the barren plateaus problem,[32] and so on. It is still unclear whether quantum chemical calculations can be accelerated from the classical computation by using VQE.

On the other hand, the quantum phase estimation (QPE) algorithm[33,34] has been paid considerable attention, because it could achieve exponential acceleration of the full-CI calculations from the classical counterpart[35] if an initial wave function that closely matches the solution can be efficiently prepared.[36] In addition, QPE is suitable for the study of excited states because we can calculate both the ground and the excited states on the same footing, by changing the input wave function. Unfortunately, the quantum circuits used in the QPE-based quantum chemical calculations are too deep to execute on NISQ devices, and only handful calculations using actual quantum computer hardware have been reported.[37–43] Currently, QPE is mainly studied using state-vector simulators those run on a classical computer. Since the computational cost for numerical simulations on a classical computer scale exponentially with the number of qubits, it is difficult to handle large molecules and currently the scope of application is still limited to small molecules such as $H_2$, $H_2O$, $CF_2$, $CFCl$, $HCHO$, $CH_4$, and 1,3-butadiene[37–51] Note that computational cost estimations and assessments for QPE have been reported.[52–55]

In anticipation of chemical industrial applications of quantum computers, it is very important to establish the computational workflow of QPE-based quantum chemical calculations that is applicable to larger and more complicated systems. For larger molecules, the appropriate selection of the active space is essential for the time being, and for excited-state calculations, an adequate initial wave-function preparation is necessary. In order to perform various numerical calculations on large molecules to search for appropriate computational conditions, speeding up numerical simulations is also an important issue.

In this work, we demonstrated a workflow for calculating electronic states using the QPE algorithm on a simulator. We implemented the algorithm with General-Purpose computing on Graphics Processing Units (GPGPUs) in conjunction with NVIDIA cuQuantum[56] to accelerate the numerical simulations. We calculated the electronic ground and the first spin-singlet π–π* excited states of benzene, chlorobenzene, and nitrobenzene, focusing on the substituent effect on the excitation energies. These chemicals are important because they are the simplest molecules containing aromatic ring and vital for their diverse roles in manufacturing, such as intermediates in producing everyday products like plastics, dyes, and pharmaceuticals.

## 2. Theory

In this section we explain the theoretical backgrounds along with our procedure. The whole procedure is summarised in Fig. 1. The computations consist of three steps: low-level quantum chemical calculations on a classical computer, pre-processing for quantum computation, and the quantum computation. As described above, we have to take care about possible obstacles in the calculation of large molecules, such as the appropriate selection of the molecular orbitals for the wave function expansion and the preparation of the initial wave function for the excited state calculation. In this work, we adopted the pseudo-natural orbitals constructed from the second order Møller–Plesset (MP2) calculation[24] as the reference molecular orbitals, and the CISD calculation within the active space to find the main configurations of the excited state wave function, as discussed in the following subsections in detail.

### 2-1. Low-level quantum chemical calculations on a classical computer

The QPE algorithm enables us to compute the full-CI energy with qubits of the same number of spin-orbitals for wave function storage and additional qubits for energy readout (ancilla qubits). In this work, we introduced the active space approximation to make the problem tractable with the state-vector simulator due to the size of the molecules under study. Under the active space approximation, the QPE returns the energy eigenvalues at the CAS-CI level of theory. Thus, generations of molecular orbitals that are suitable bases and selection of appropriate active orbitals are crucial for reducing the dimensionality of the problem and obtain accurate energies in QPE. In this study, we constructed the pseudo-natural orbitals from the MP2 calculations and used them for the CAS-CI calculations. Since the natural orbitals are obtained by diagonalising the one-body density matrix and give the occupation numbers as their eigenvalues, it is convenient to use the (pseudo-)natural orbitals to find important molecular orbitals with respect to the ground state wave function. This strategy is often used in the CASSCF and the CAS-CI calculations[57]. In the context of quantum computing, it has been reported that the Trotter–Suzuki decomposition error discussed below can be reduced by using the natural orbitals instead of the Hartree–Fock (HF) canonical orbitals.[58]

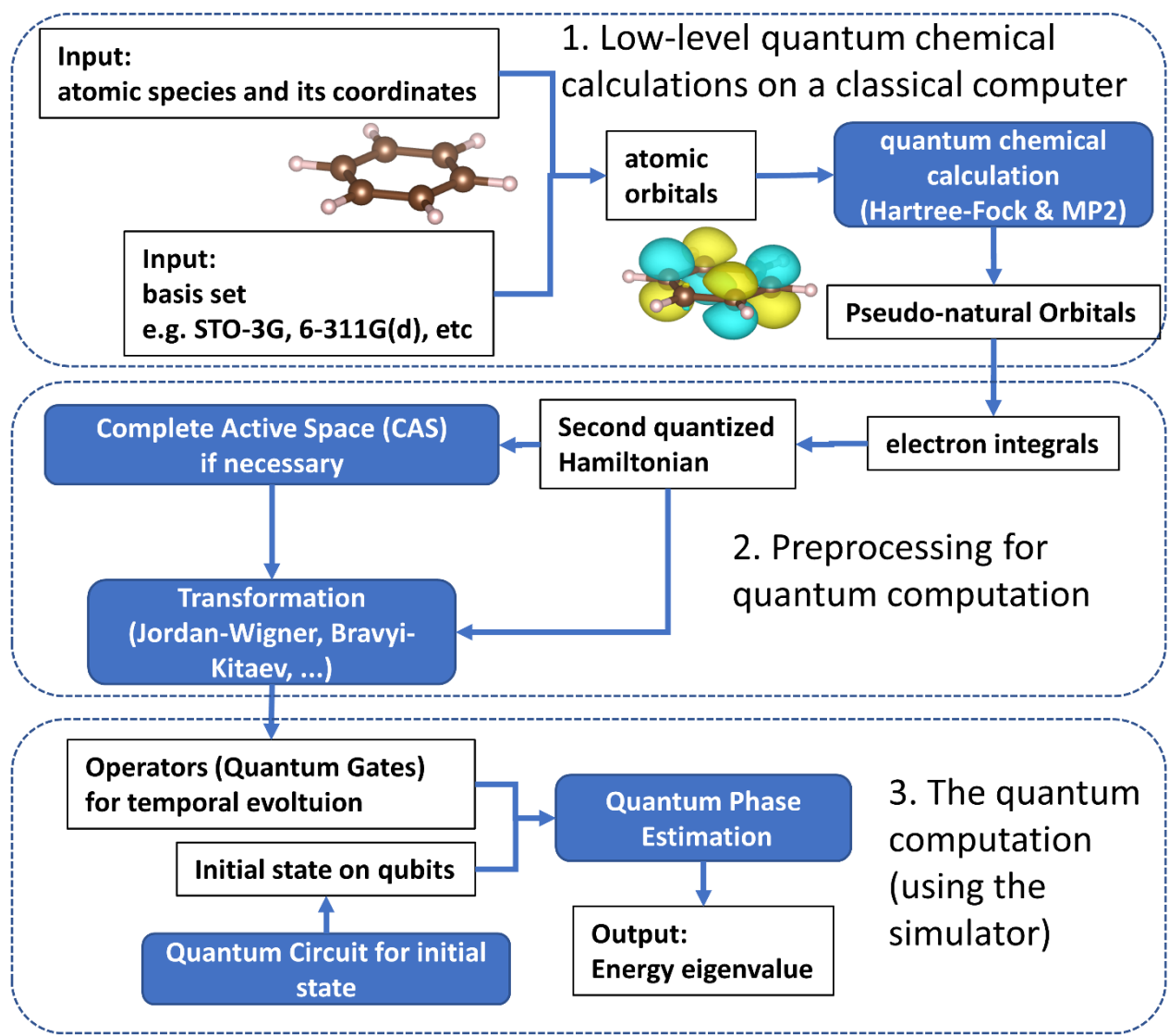

Fig. 1 The workflow of our quantum chemical calculation on quantum computer. It consists of three parts. (1) Low-level quantum chemical calculations on a classical computer, which aims to obtain the information of molecular orbitals. (2) Pre-processing for quantum computation, which is to transform the above obtained molecular orbitals into quantum gates and qubits. (3) The Quantum computations (using the simulator, or if possible, using a quantum computer).

As pointed out in the Introduction section, QPE requires an approximate wave function of the target electronic state. For the electronic ground state calculations of closed-shell molecules in the vicinity of their equilibrium geometry, the HF wave function is often used as the input wave function (although this strategy does not work for strongly correlated systems and for very large molecules such as proteins). In this study we used the HF-like single configurational wave function in the MP2 pseudo-natural orbital basis as the approximate wave function for the ground state, as suggested by reference. For the excited state calculations, we first performed the CISD calculation in the active space to find the major configurations of the excited state, and the approximate wave function is constructed accordingly by considering the main configurations.

**2-2. Preprocessing for quantum computation**

The next step is to obtain the expressions of Hamiltonian and wave functions in the quantum computing language. With the molecular orbitals obtained above, we can construct the second-quantised Hamiltonian as follows[24]:

$$\mathcal{H} = \sum_{p,q} h_{pq}\, a_p^\dagger a_q + \frac{1}{2}\sum_{p,q,r,s} h_{pqrs}\, a_p^\dagger a_q^\dagger a_s a_r \quad (1)$$

Here, $h_{pq}$ and $h_{pqrs}$ are one- and two- electron integrals, respectively, calculated using the pseudo-natural orbitals. $a_p^\dagger$ and $a_p$ are creation and annihilation operators, respectively, acting on the *p*-th spin-orbital. Indices *p*, *q*, *r*, and *s* run over the active orbitals being selected.

Once we get the second-quantised Hamiltonian in eqn (1), we can obtain the expression on quantum computers with the help of appropriate fermion–qubit encoding methods which transforms fermionic creation and annihilation operators to qubit operators comprised of Pauli operators as given in eqn (2).

$$\mathcal{H} = \sum_{j=1}^{J} c_j P_j \quad (2)$$

Here, $P_j$ is a direct product of Pauli operators as in eqn (3) called as a Pauli string, and $c_j$ is the corresponding coefficients calculated from $h_{pq}$ and $h_{pqrs}$. $J$ is the number of Pauli strings in the qubit Hamiltonian.

$$P_j = \sigma_1 \otimes \sigma_2 \otimes \cdots \otimes \sigma_N,\ \sigma \in \{I, X, Y, Z\} \quad (3)$$

In this report we chose the Jordan–Wigner transformation (JWT)[59] for the fermion–qubit encoding. In the JWT we associate one spin-orbital with one qubit. We assign $|1\rangle$ for a qubit if the corresponding spin-orbital is occupied by an electron, otherwise $|0\rangle$. A Slater determinant of the system is expressed as the product state of these qubits. For example, consider about the HF configuration in (6e, 6o) active apace, where (*N*e, *M*o) represent *N* active electrons in *M* molecular orbitals. Under the JWT the HF state is given as $|111111000000\rangle$. Once the qubit Hamiltonian is generated, one can easily construct the quantum circuit corresponding to the time evolution operator $U = \exp(-i\mathcal{H}t)$ using the reported procedure.[60]

## 2-3. Quantum Computation

The last step is performing the quantum chemical calculations on a quantum computer using the QPE algorithm. Detailed introduction of the QPE can be found elsewhere.[61] There are several derivatives of QPE, e.g., Kitaev's QPE,[62] *N*-qubit QPE as a textbook implementation,[61] iterative QPE (IQPE),[63,64] and Bayesian QPE.[39,65,66] As the advanced techniques for the QPE, quantum phase difference estimation (QPDE) algorithms for the direct calculation of energy differences[43,45–48,51] and quantum multiphase estimation methods,[67,68] etc, have been reported. Among them we selected the IQPE algorithm because it can be implemented with only one ancilla qubit as shown below, and it returns the eigenvalue of Hamiltonian with an approximate wave function as the input.

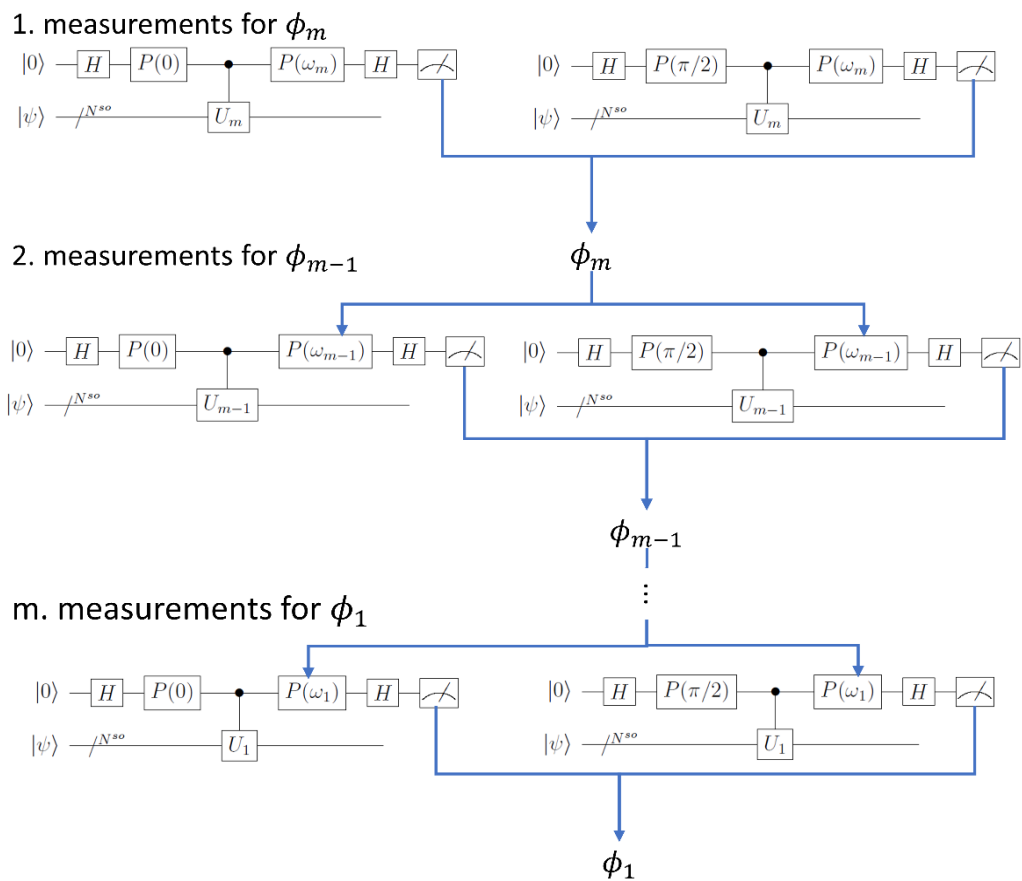

**Fig. 2** Quantum Circuits for the IQPE algorithm. First qubit expressed as $|0\rangle$ is the ancilla qubit, and other $N^{so}$ qubits (represented as $|\psi\rangle$ in the figure) are used for wave function mapping. $H$ and $P(\omega)$ are Hadamard and phase shift gates defined in eqn (5) and (6), respectively, in the main text. Measurements are first performed for the least important digit of the phase ("1. measurements for $\phi_m$"). Based on the result, we set the angle for the phase shift, and the measurements for the second least important digit are performed ("2. measurements for $\phi_{m-1}$"). This process is repeated until the phase is obtained for all digits ("m. measurements for $\phi_1$").

Below we show the algorithms of IQPE. In the IQPE algorithms the energy eigenvalue $E$ can be obtained by simulating the time evolution of wave function and extracting the amount of the phase shift $\phi$ ($0 \leq \phi < 1$) caused by the time evolution, which is related to the energy $E$ as in eqn (4).

$$U|\psi\rangle = e^{-i\mathcal{H}t}|\psi\rangle = e^{-iEt}|\psi\rangle = e^{i2\pi\phi}|\psi\rangle \quad (4)$$

The quantum circuit for the IQPE algorithm is given in Fig. 2. Here, $H$ and $P(\theta)$ are Hadamard and phase shift gates defined in eqn (5) and (6), respectively.

$$H = \tfrac{1}{\sqrt{2}}\begin{pmatrix} 1 & 1 \\ 1 & -1 \end{pmatrix} \quad (5)$$

$$P(\theta) = \begin{pmatrix} 1 & 0 \\ 0 & e^{i\theta} \end{pmatrix} \quad (6)$$

In the IQPE, we perform the quantum operations in an iterative manner to sequentially project the initial state onto the eigenfunction. Here we adopt the algorithm proposed in the reference,[69] and our implementation is based on the procedure in the literature,[70] which is arranged to fit our case due to the limitation in computational resource.

Let us decompose $\phi$ as follows:

$$\phi = \sum_{k=1}^{m} 2^{-k}\phi_k + \delta 2^{-m} \quad (7)$$

Here $\phi_k$ is the $k$-th digit of the phase in the fractional binary to be determined from the measurement of an ancilla qubit, $m$ is the number of digits to be calculated, and $\delta$ is the error such that $0 \leq |\delta| < 1$.

First, we determine the least important digit $\phi_m$. To do this, we performed two separate quantum circuit executions in Fig. 2 with $U_m = \exp(-i\mathcal{H}t)^{2^{m-1}}$ with the sets of rotational angles $(\theta, \omega_m) = (0, 0)$ and $(\pi/2, 0)$, respectively. After the operations, the expected phase of the state, $\phi'$, is given as

$$\phi' = \sum_{k=1}^{m-1} 2^{m-1-k}\phi_k + 2^{-1}\phi_m + \delta 2^{-1} \quad (8)$$

The first term, contributions from $\phi_k$s ($0 \leq k < m$), can be ignored because these are integers and the resulting change in the state is expressed as multiplying by 1; i.e. meaning that no effects on the measurement result. Therefore, we can focus on the remaining terms. By the effect of these terms, the input state $|0\rangle \otimes |\psi\rangle$ is transformed into the state given in eqn (9):

$$\tfrac{1+e^{i\theta}e^{i\pi\phi_m}e^{i\pi\delta}}{2}|0\rangle \otimes |\psi\rangle + \tfrac{1-e^{i\theta}e^{i\pi\phi_m}e^{i\pi\delta}}{2}|1\rangle \otimes |\psi\rangle \quad (9)$$

The probability to obtain the $|0\rangle$ state in the measurement of an ancilla qubit can be calculated as follows.

$$Prob_0(\theta) = \tfrac{1+\cos(\theta+\phi_m+\delta)}{2} \quad (10)$$

Since we make operations for $\theta = 0$ and $\pi/2$, we have values of $Prob_0(0)$ and $Prob_0(\pi/2)$. With these values we can get

$$\cos(\phi_m + \delta) = \tfrac{2Prob_0(0)-1}{2} \quad (11)$$

$$\sin(\phi_m + \delta) = \tfrac{2Prob_0(\pi/2)-1}{2} \quad (12)$$

This means that we can determine $\phi_m + \delta$. The accuracy is dependent on the number of samplings.

Next is to determine $\phi_k$ in order from $k = m - 1$ to 1. This is done by setting $U$ for $U_k = \exp(-i\mathcal{H}t)^{2^{k-1}}$ and $\omega_k = -\sum_{i=1}^{m-k-1} 2^{-i}\phi_{k+i} + 2^{m-k}(\phi_m + \delta)$, with setting the output wave function in the previous iteration as the input. By measuring ancilla qubit with $\theta = 0$ we obtain $\phi_k$ such that

$$\cos\phi_k = \tfrac{2Prob_0(0)-1}{2} \quad (13)$$

By sequentially repeating this procedure we can determine all the digits.

**2-4. Computational details**

As mentioned above, speeding up the numerical simulation is essential to perform the QPE simulations of large molecules to construct a practical workflow. Before tackling the QPE simulations of benzene and its derivatives, we performed the IQPE quantum circuit simulations of the LiH molecule on CPUs (from 1 to 48) and on a GPGPU, to compare the simulation time. Molecular geometry of LiH was optimised at the B3LYP[71–73]/6-31+G(d) level of theory using Gaussian16,[74] and it was used for the IQPE single point calculation. Cartesian coordinates of the optimised geometry are provided in Table S.1.1 of the ESI†. We examined different sizes of the problem by changing the number of active orbitals from six to ten, e.g., the number of qubits is 13 when we include 6 natural orbitals in the calculation.

In order to validate the effectiveness of the proposed workflow, we performed the IQPE simulations for benzene, chlorobenzene, and nitrobenzene, with $m = 12$. Geometry optimizations were done at the B3LYP/6-311G(d) method. Cartesian coordinates of the optimised geometries are provided in Tables S.1.2, 1.3, 1.4 of the ESI†. We calculated the total energies of the ground state and the first spin-singlet π–π* excited state to estimate the excitation energies. We compared them with the experimental values[75–77] and the CAS-CI excitation energies as the reference. We selected six valence π and π* orbitals in benzene ring and the $2p_z$-type orbitals of substituents those participating the π-conjugation as the active orbitals. The active space is (6e, 6o), (8e, 7o), and (10e, 9o) for benzene, chlorobenzene, and nitrobenzene, respectively. All the active orbitals are illustrated in Fig. S.2.1, S.2.2, S.2.3 in the ESI†.

In this study we used the HF-like single configurational wave function as the input wave function for the electronic ground state. For the excited state calculations of benzene and chlorobenzene, we first perform the CISD calculations within the active space and selected two major configurations (HONO−1 → LUNO) and (HONO → LUNO+1). Here, HONO and LUNO stand for the highest occupied natural orbital and the lowest unoccupied natural orbital, respectively. For nitrobenzene, we also included (HONO → LUNO) excited configuration and construct the 3-configurational wave

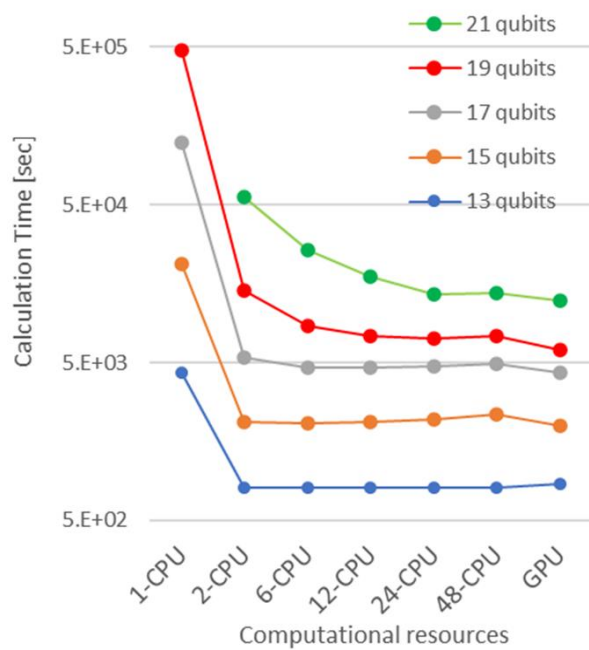

**Fig. 3** The comparison of the calculation time of numerical simulations with multi-CPUs or a single-GPGPU

## 3. Results and Discussions

### 3-1. The speed up of the Simulator

First, we investigated the calculation time dependence on the problem size i.e., the number of qubits. The results are summarised in Fig. 3. The execution time decreases significantly by using multi-CPUs. However, the speedup effect is saturated with a few CPUs, beyond which further improvements cannot be observed. It is noteworthy that the computation time with a single-GPGPU is generally the lowest except for the 13-qubit simulations, presumably due to a decrease in the percentage of total computation time spent on quantum circuit simulations. Comparing with single-CPU case, the computation time is reduced to 1/5 and 1/80 in case of 13 and 19 qubits, respectively, in single-GPGPU. As a result, we can deduce that IQPE simulations can be carried out more efficiently with single-GPGPU than with multiple CPUs. Based on this finding, all subsequent computations were performed exclusively using GPGPU.

### 3-2. Calculations for Benzene, Chlorobenzene, and Nitrobenzene

Secondly, we investigated the π–π* excitation energies of benzene, chlorobenzene, and nitrobenzene. The results are summarised in Table 1. The computational time for the single state (the ground or excited state) of these molecules are 19, 95, 206 hours, respectively.

As clearly seen in Table 1, both the CAS-CI and the IQPE reproduced the experimental tendency of the π–π* excitation energies. We observe discrepancies in the obtained values with the experimental ones, due to insufficient consideration of dynamical electron correlation effects. In our CAS-CI and IQPE calculations we only consider valence π electrons/orbitals in the active space, which can lead to the over-estimation of excitation energies.[31] We believe that this discrepancy can be reduced by using more extended active space or combining our method with some perturbation methods on quantum computer that are recently proposed,[85,86] which is out of scope of this work.

functions as the input. The ratios of the CI coefficients were kept the same as the results of the CISD calculations. We changed the number of excited configurations included in the approximate wave function in a flexible way, considering the trade-off between the number of excited configurations and the depth of the quantum circuit for the state preparation. By using this approach, we can generate the approximate excited state wave function with the overlap squared value $|\langle\Psi_{approx}|\Psi_{CAS-CI}\rangle|^2$ = 0.81, 0.81, and 0.85, for benzene, chlorobenzene, and nitrobenzene, respectively. The quantum circuits used for the excited state wave function generation were given as Fig. S.4.1, 4.2, 4.3 in the ESI†.

In the IQPE algorithm we must simulate the time evolution of a wave function. In this study we adopted the second-order Trotter–Suzuki decomposition[78,79] given in eqn (14) to construct the quantum circuit for the time evolution operator.

$$e^{-\sum_{j=1}^{J} i\omega_j P_j t} \approx \left[\prod_{j=1}^{J} e^{-\frac{i\omega_j P_j t}{2\mathcal{M}}} \prod_{j=J}^{1} e^{-\frac{i\omega_j P_j t}{2\mathcal{M}}}\right]^{\mathcal{M}} \quad (14)$$

Here, $\mathcal{M}$ is the number of slices in the Trotter–Suzuki decomposition. In this study, the evolution time length $t$ in $U = \exp(-i\mathcal{H}t)$ and $\mathcal{M}$ were set to be 1.0 and 5, respectively.

All the quantum chemical calculations except for geometry optimizations were performed with PySCF.[80] We developed a Python3 program for the IQPE quantum circuit simulations, using Cirq (ver. 1.0.0)[81] and OpenFermion (ver. 1.6.0).[82] For the GPGPU-based simulations, we used qsimcirq[83] build locally. The operations were performed on Intel(R) Xeon(R) Platinum 8168 CPU @ 2.70GHz with 96 CPU cores and Tesla V-100 GPGPU unit (NVIDIA), deployed in NVIDIA DGX-2 Super-computer in our private environment.[84]

Table 1. The calculated and experimental π–π* excitation energy of benzene, chlorobenzene, and nitrobenzene, in units of eV.

| Molecule | Experimental | with CAS-CI | with IQPE |
|---|---|---|---|
| Benzene | 4.900 | 6.091 | 6.092 |
| Chlorobenzene | 4.720 | 5.998 | 6.008 |
| Nitrobenzene | 4.380 | 5.951 | 5.925 |

It is noteworthy that the IQPE reproduced the CAS-CI excitation energies within 30 meV of error. This error can be explained by the truncation error on the phase readout with finite digits and the error arising from Trotter–Suzuki decomposition. In the present calculations, the phase values were determined up to 12 digits, so $\delta\phi$, the error in phase, is in the order of $2^{-12}$. According to eqn (4) and (7) with $t$ = 1, the estimated error of the energy, $\delta E$, can be calculated as follows:

$$\Delta E = \frac{2\pi\Delta\phi}{t} \sim 41.7 \text{ meV} \quad (15)$$

The value is comparable to chemical precision (1.0 kcal mol$^{-1}$ ~ 43 meV). The error in the excitation energy can be twice as large as in eqn (15), because of the additive nature of the error in the energy difference. The observed energy discrepancies between the CAS-CI and the IQPE are smaller than this value, suggesting that the errors in the ground and excited state energies cancel out to some extent.

Note that in the IQPE algorithm we did not consider configuration interactions explicitly. We just enumerated one- and two- electron integrals and transformed them into quantum gates. In the IQPE quantum circuit for the determination of kth digit, the measurement of the ancilla qubit projects the input wave function to a set of eigenstates whose eigenphases $\phi$ of the kth digit match the measurement result. The ability of the IQPE algorithm giving the eigenvalue and corresponding eigenstate is clearly demonstrated in our numerical simulations.

## Conclusions

In summary, we demonstrated the workflow for calculating electronic states of benzene and its mono-substituted derivatives using the IQPE algorithm on a simulator. We implemented the algorithm with single-GPGPU, observing ×80 speedup for 19-qubit simulations. By generating the pseudo-natural orbital at the MP2 level of theory, we can easily construct the active space suitable for the computations. In the π–π* excited states calculations of benzene and its derivatives, we adopted an approach of performing the CISD calculation within the active space to identify the major configurations of the excited state wave function, to prepare approximate wave function of the excited state used as the input in IQPE quantum circuit. The π–π* excitation energies calculated at the IQPE quantum circuit simulations agreed with the CAS-CI excitation energies with less than 30 meV of deviations, achieving the chemical precision (1.0 kcal mol$^{-1}$ ~ 43 meV). To tackle with larger molecules with more active orbitals, computational cost reduction by adopting qubit tapering techniques[87,88] and quantum circuit optimizations[49,89] must be necessary. The studies along this line are underway.

## Author Contributions

Y. Ino, H. Yuzawa, Y. Minato and K. Sugisaki planned and conducted the project. Y. Ino and K. Sugisaki carried out the quantum chemical calculations. Y. Ino, M. Yonekawa, H. Yuzawa and K. Sugisaki developed the quantum circuit simulation programs and performed numerical simulations. All the authors discussed the results and wrote the paper.

## Data availability

The authors confirm that the data supporting the findings of this study are available within the article [and/or its supplementary materials].

## Conflicts of interest

There are no conflicts to declare.

## Acknowledgements

This work has been partially supported from the Center of Innovations for Sustainable Quantum AI (Grant No. JPMJPF2221) from JST, Japan. Y.I., M.Y. and H.Y. acknowledge the support of S. Sugimoto at the Imaging & Informatics Laboratories of ICT Headquarters in FUJIFILM Corporation. Y.I., M.Y. and H.Y. also thank T. Yoshioka and H. Ishikawa for helpful discussions. K.S. acknowledges the support from Quantum Leap Flagship Program (JPMXS0120319794) from the MEXT, Japan, and KAKENHI Transformative Research Area B (23H03819) and Scientific Research C (21K03407) from JSPS, Japan.

## References

1  E. R. MacQuarrie, C. Simon, S. Simmons and E. Maine, The emerging commercial landscape of quantum computing, *Nat. Rev. Phys.*, 2020, **2,** 11, 596.
2  M. Motta and J. E. Rice, Emerging quantum computing algorithms for quantum chemistry, *Wiley Interdiscip. Rev. Comput. Mol. Sci.*, 2022, **12,** e1580.
3  H. P. Cheng, E. Deumens, J. K. Freericks, C. Li and B. A. Sanders, Application of quantum computing to biochemical systems: a look to the future, *Front. Chem.*, 2020, **8,** 587143.
4  C. Outeiral, M. Strahm, J. Shi, G. M. Morris, S. C. Benjamin and C. M. Deane, The prospects of quantum computing in computational molecular biology, *Wiley Interdiscip. Rev. Comput. Mol. Sci.*, 2021, **11,** e1481.
5  L. Bassman, M. Urbanek, M. Metcalf, J. Carter, A. F. Kemper and W. A. De Jong, Simulating quantum materials with digital quantum computers, *Quantum Sci. Technol.*, 2021, **6,** 043002.
6  F. Tacchino, A. Chiesa, S. Carretta and D. Gerace, Quantum computers as universal quantum simulators: state-of-the-art and perspectives, *Adv. Quantum Technol.,* 2020, **3,** 1900052.
7  D. Herman, C. Googin, X. Liu, Y. Sun, A. Galda, I. Safro, M. Pistoia and Y. Alexeev, Quantum computing for finance, *Nat. Rev. Phys.*, 2023, **5,** 450.
8  A. Perdomo-Ortiz, M. Benedetti, J. Realpe-Gómez and R. Biswas, Opportunities and challenges for quantum-assisted machine learning in near-term quantum computers, *Quantum Sci. Technol.*, 2018, **3,** 030502.
9  T. M. Fernandez-Carames and P. Fraga-Lamas, Towards post-quantum blockchain: a review on blockchain cryptography resistant to quantum computing attacks, *IEEE Access*, 2020, **8,** 21091–21116.
10  National Academies of Sciences, Engineering, and Medicine; Division on Engineering and Physical Sciences; Computer Science and Telecommunications Board; Intelligence Community Studies Board; Committee on Technical Assessment of the Feasibility and Implications of Quantum Computing; Emily Grumbling and Mark Horowitz, Editors., *Quantum Computing: Progress and Prospects*, The National Academies Press, Washington DC, 2019.
11  Y. Kim, A. Eddins, S. Anand, K. X. Wei, E. van den Berg, S. Rosenblatt, H. Nayfeh, Y. Wu, M. Zaletel, K. Temme and A. Kandala, Evidence for the utility of quantum computing before fault tolerance, *Nature,* 2023, 618, 500–505.
12  J. Robledo-Monero, M. Motta, H. Haas, A. Javadi-Abhari, P. Jurcevic, W. Kirby, S. Martiel, K. Sharma, S. Sharma, T. Shirakawa, I. Sitdikov, R.-Y. Sun, K. J. Sung, M. Takita, M. C.

Tran, S. Yunoki and A. Mezzacapo, Chemistry beyond exact solutions on a quantum-centric supercomputer, *arXiv*, 2024, preprint, arXiv:2405.05068. https://doi.org/10.48550/arXiv.2405.05068

13  D. Bluvstein, S. J. Evered, A. A. Geim, S. H. Li, H. Zhou, T. Manovitz, S. Ebadi, M. Cain, M. Kalinowski, D. Hangleiter, J. P. B. Ataides, N. Maskara, I. Cong, X. Gao, P. S. Rodriguez, T. Karolyshyn, G. Semeghini, M. J. Gullans, M. Greiner, V. Vuletić and M. D. Lukin, Logical quantum processor based on reconfigurable atom arrays, *Nature,* 2023, **626**, 58.

14  Y. Cao, J. Romero, J. P. Olson, M. Degroote, P. D. Johnson, M. Kieferová, I. D. Kivlichan, T. Menke, B. Peropadre, N. P. D. Sawaya, S. Sim, L. Veis and A. Aspuru-Guzik, Quantum chemistry in the age of quantum computing, *Chem. Rev.*, 2019, **119,** 10856–10915.

15  B. Bauer, S. Bravyi, M. Motta and G. K.-L. Chan, Quantum algorithms for quantum chemistry and quantum materials science, *Chem. Rev.*, 2020, **120**, 12685–12717.

16  H. P. Paudel, M. Syamlal, S. E. Crawford, Y.-L. Lee, R. A. Shugayev, P. Lu, P. R. Ohodnicki, D. Mollot, and Y. Duan, Quantum computing and simulations for energy applications: review and perspective, *ACS Eng. Au*, 2022, **2**, 151–196.

17  N. S. Blunt, J. Camps, O. Crawford, R. Izsák, S. Leontica, A. Mirani, A. E. Moylett, S. A. Scivier, C. Sünderhauf, P. Schopf, J. M. Taylor, and N. Holzmann, Perspective on the current state-of-the-art of quantum computing for drug discovery Applications, *J. Chem. Theory Comput*., 2022, **18**, 7001–7023.

18  F. Häse, L. M. Roch, P. Friederich and A. Aspuru-Guzik, Designing and understanding light-harvesting devices with machine learning, *Nat. Commun.*, 2020, **11**, 4587.

19  Y. Sumiya, Y. Harabuchi, Y. Nagata and S. Maeda, Quantum chemical calculations to trace back reaction paths for the prediction of reactants, *JACS Au* 2022, **2,** 1181–1188.

20  B. Craig, C.-K. Skylaris, T. Schoetz and C. P. de León, A computational chemistry approach to modelling conducting polymers in ionic liquids for next generation batteries, *Energy Rep.*, 2020, **6**, 198–208.

21  R. M. Borges, S. M. Colby, S. Das, A. S. Edison, O. Fiehn, T. Kind, J. Lee, A. T. Merrill, K. M. Merz Jr., T. O. Metz, J. R. Nunez, D. J. Tantillo, L.-P. Wang, S. Wang and R. S. Renslow, Quantum chemistry calculations for metabolomics, *Chem. Rev.,* 2021, **121**, 5633–5670.

22  W. Kohn and L. J. Sham, Self-consistent equations Including exchange and correlation effects, Phys. Rev. 1965, **140**, A1133–A1138.

23  A. J. Cohen, P. Mori-Sánchez and W. Yang, Challenges for density functional theory, Chem. Rev. 2012, **112**, 289–320.

24  T. Helgaker, P. Jørgensen and J. Olsen, Molecular Electronic Structure Theory, John Wiley & Sons, Inc, Chichester, 2000.

25  B. G. Levine, A. S. Durden, M. P. Esch, F. Liang and Y. Shu, CAS without SCF—Why to use CASCI and where to get the orbitals, *J. Chem. Phys.*, 2021, **154,** 090902.

26  S. Pijeau and E. G. Hohenstein, Improved complete active space configuration interaction energies with a simple correction from density functional theory, *J. Chem. Theory Comput.*, 2017, **13**, 1130–1146.

27  J. Tilly, H. Chen, S. Cao, D. Picozzi, K. Setia, Y. Li, E. Grant, L. Wossnig, I. Rungger, G. H. Booth and J. Tennyson, The variational quantum eigensolver: a review of methods and best practices, *Phys. Rep.*, 2022, **986**, 1–128.

28  A. Kandala, A. Mezzacapo, K. Temme, M. Takita, M. Brink, J. M. Chow and J. M. Gambetta, Hardware-efficient variational quantum eigensolver for small molecules and quantum magnets, *Nature,* 2017, **549**, 242–246.

29  A. Peruzzo, J. McClean, P. Shadbolt, M.-H. Yung, X.-Q. Zhou, P. J. Love, A. Aspuru-Guzik and J. L. O'Brien, A variational eigenvalue solver on a photonic quantum processor, *Nat. Commun.*, 2014, **5**, 4213.

30  J. F. Gonthier, M. D. Radin, C. Buda, E. J. Doskocil, C. M. Abuan, and J. Romero, Measurements as a roadblock to near-term practical quantum advantage in chemistry: resource analysis, *Phys. Rev. Res.*, 2022, **4**, 033154.

31  L. Bittel and M. Kliesch, Training variational quantum algorithms is NP-hard, *Phys. Rev. Lett.* 2021, **127**, 120502.

32  J. R. McClean, S. Boixo, V. N. Smelyanskiy, R. Babbush and H. Neven, Barren plateaus in quantum neural network training landscapes, *Nat. Commun.*, 2018, **9**, 4812.

33  D. S. Abrams and S. Lloyd, Simulation of many-body Fermi systems on a universal quantum computer, *Phys. Rev. Lett.*, 1997, **79**, 2586–2589.

34  D. S. Abrams and S. Lloyd, Quantum algorithm providing exponential speed increase for finding eigenvalues and eigenvectors, *Phys. Rev. Lett.*, 1999, **83**, 5162–5165.

35  A. Aspuru-Guzik, A. D. Dutoi, P. J. Love and M. Head-Gordon, Simulated quantum computation of molecular energies, *Science*, 2005, **309,** 1704–1707.

36  S. Lee, J. Lee, H. Zhai, Y. Tong, A. M. Dalzell, A. Kumar, P. Helms, J. Gray, Z.-H. Cui, W. Liu, M. Kastoryano, R. Babbush, J. Preskill, D. R. Reichman, E. T. Campbell, E. F. Valeev, L. Lin and G. K.-L. Chan, Evaluating the evidence for exponential quantum advantage in ground-state quantum chemistry, *Nat. Commun.*, 2023, **14**, 1952.

37  J. Du, N. Xu, X. Peng, P. Wang, S. Wu and D. Lu, NMR implementation of a molecular hydrogen quantum simulation with adiabatic state preparation, *Phys. Rev. Lett.*, 2010, **104**, 030502.

38  Y. Wang, F. Dolde, J. Biamonte, R. Babbush, V. Bergholm, S. Yang, I. Jakobi, P. Neumann, A. Aspuru-Guzik, J. D. Whitfield and J. Wrachtrup, Quantum simulation of helium hydride cation in a solid-state spin register, *ACS Nano*, 2015, **9**, 7769–7774.

39  K. Yamamoto, S. Duffield, Y. Kikuchi and D. Muñoz Ramo, Demonstrating Bayesian quantum phase estimation with quantum error detection, *Phys. Rev. Res.*, 2024, **6**, 013211.

40  B. P. Lanyon, J. D. Whitfield, G. G. Gillett, M. E. Goggin, M. P. Almeida, I. Kassal, J. D. Biamonte, M. Mohseni, B. J. Powell, M. Barbieri, A. Aspuru-Guzik and A. G. White, Towards quantum chemistry on a quantum computer, *Nat. Chem.*, 2010, **2**, 106–111.

41  P. J. J. O'Malley, R. Babbush, I. D. Kivlichan, J. Romero, J. R. McClean, R. Barends, J. Kelly, P. Roushan, A. Tranter, N. Ding, B. Campbell, Y. Chen, Z. Chen, B. Chiaro, A. Dunsworth, A. G. Fowler, E. Jeffrey, E. Lucero, A. Megrant, J. Y. Mutus, M. Neeley, C. Neill, C. Quintana, D. Sank, A. Vainsencher, J. Wenner, T. C. White, P. V. Coveney, P. J. Love, H. Neven, A. Aspuru-Guzik and J. M. Martinis, Scalable quantum simulation of molecular energies, *Phys. Rev. X*, 2016, **6**, 031007.

42  N. S. Blunt, L. Caune, R. Iszák, E. T. Campbell and N. Holzmann, Statistical phase estimation and error mitigation on a superconducting quantum processor, *PRX Quantum*, 2023, **4**, 040341.

43  S. Kanno, K. Sugisaki, H. Nakamura, H. Yamauchi, R. Sakuma, T. Kobayashi, Q. Gao, N. Yamamoto, Tensor-based quantum phase difference estimation for large-scale demonstration, arXiv, 2024, preprint, arXiv:2408.04946, https://doi.org/10.48550/arXiv.2408.04946

44  C. Kang, N. P. Bauman, S. Krishnamoorthy and K. Kowalski, Optimized quantum phase estimation for simulating electronic states in various energy regimes, *J. Chem. Theory Comput.*, 2022, **18**, 6567–6576.

45  K. Sugisaki, C. Sakai, K. Toyota, K. Sato, D. Shiomi and T. Takui, Bayesian phase difference estimation: a general quantum algorithm for the direct calculation of energy gaps, *Phys. Chem. Chem. Phys.*, 2021, **23**, 20152–20162.

46  K. Sugisaki, C. Sakai, K. Toyota, K. Sato, D. Shiomi and T. Takui, Quantum algorithm for full configuration interaction

calculations without controlled time evolutions, *J. Phys. Chem. Lett.*, 2021, **12**, 11085–11089.
47 K. Sugisaki, H. Wakimoto, K. Toyota, K. Sato, D. Shiomi and T. Takui, Quantum algorithm for numerical energy gradient calculations at the full configuration interaction level of theory, *J. Phys. Chem. Lett.*, 2022, **13**, 11105–11111.
48 K. Sugisaki, V. S. Prasannaa, S. Ohshima, T. Katagiri, Y. Mochizuki, K. Sahoo and B. P. Das, Bayesian phase difference estimation algorithm for direct calculation of fine structure splitting: accelerated simulation of relativistic and quantum many-body effects, *Electron. Struct.*, 2023, **5**, 035006.
49 N. Baskaran, A. S. Rawat, A. Jayashankar, D. Chakravarti, K. Sugisaki, S. Roy, S. B. Mandal, D. Mukherjee and V. S. Prasannaa, Adapting the Harrow-Hassidim-Lloyd algorithm to quantum many-body theory, *Phys. Rev. Res.* 2023, **5**, 043113.
50 N. P. Bauman, H. Liu, E. J. Bylaska, S. Krishnamoorthy, G. H. Low, C. E. Granade, N. Wiebe, N. A. Baker, B. Peng, M. Roetteler, M. Troyer and K. Kowalski, Toward quantum computing for high-energy excited states in molecular systems: quantum phase estimations of core-level states, *J. Chem. Theory Comput.*, 2021, **17**, 201–210.
51 K. Sugisaki, Projective measurement-based quantum phase difference estimation algorithm for the direct computation of eigenenergy differences on a quantum computer, *J. Chem. Theory Comput.*, 2023, **19,** 7617–7625.
52 A. Delgado, P. A. M. Casares, R. dos Reis, M. S. Zini, R. Campos, N. Cruz-Hernández, A.-C. Voigt, A. Lowe, S. Jahangiri, M. A. Martin-Delgado, J. E. Mueller and J. M. Arrazola, Simulating key properties of lithium-ion batteries with a fault-tolerant quantum computer, *Phys. Rev. A*, 2022, **106**, 032428.
53 P. A. M. Casares, R. Campos and M. A. Martin-Delgado, TFermion: A non-Clifford gate cost assessment library of quantum phase estimation algorithms for quantum chemistry, *Quantum*, 2022, **6**, 768.
54 M. Reiher, N. Wiebe, K. M. Svore, D. Wecker and M. Troyer, Elucidating reaction mechanisms on quantum computers, *PNAS*, 2017, **114**, 7555-7560.
55 R. Babbush, C. Gidney, D. W. Berry, N. Wiebe, J. McClean, A. Paler, A. Fowler and H. Neven, Encoding electronic spectra in quantum circuits with linear T complexity, *Phys. Rev. X*, 2018, **8**, 041015.
56 H. Bayraktar, A. Charara, D. Clark, S. Cohen, T. Costa, Y.-L. L. Fang, Y. Gao, J. Guan, J. Gunnels, A. Haidar, A. Hehn, M. Hohnerbach, M. Jones, T. Lubowe, D. Lyakh, S. Morino, P. Springer, S. Stanwyck, I. Terentyev, S. Varadhan, J. Wong and T. Yamaguchi, cuQuantum SDK: a high-performance library for accelerating quantum science, in 2023 IEEE International Conference on Quantum Computing and Engineering (QCE), Bellevue, WA, USA, 2023 pp. 1050-1061.
57 Jørgensen H. J. A. Jensen, P. Jørgensen, H. Ågren and Jeppe Olsen, Second-order Møller–Plesset perturbation theory as a configuration and orbital generator in multiconfiguration self-consistent field calculations, *J. Chem. Phys.*, 1988, **88**, 3834-3839.
58 R. Babbush, J. McClean, D. Wecker, A. Aspuru-Guzik and N. Wiebe, Chemical basis of Trotter-Suzuki errors in quantum chemistry simulation, *Phys. Rev. A*, 2015, **91**, 022311.
59 P. Jordan and E. Wigner, Über das Paulische äquivalenzverbot, *Z. Phys.*, 1928, **47**, 631–651.
60 J. D. Whitfield, J. Biamonte and A. Aspuru-Guzik, Simulation of electronic structure Hamiltonians using quantum computers, *Mol. Phys.*, 2011, **109,** 735–750.
61 M. A. Nielsen and I. L. Chuang, *Quantum Computation and Quantum Information 2nd ed.*, Cambridge University Press, Cambridge, 2010.
62 S. B. Bravyi and A. Y. Kitaev, Fermionic quantum computation, *Annals Phys.* 2002, **298**, 210

63 M. Dobšíček, G. Johansson, V. Shumeiko and G. Wendin, Arbitrary accuracy iterative quantum phase estimation algorithm using a single ancillary qubit: a two-qubit benchmark, *Phys. Rev. A*, 2007, **76**, 030306.
64 D. Halder, V. S. Prasannaa, V. Agarawal and R. Maitra, Iterative quantum phase estimation with variationally prepared reference state, *Int. J. Quantum Chem.,* 2023, **123,** e27021.
65 N. Wiebe and N. Granade, Efficient Bayesian phase estimation, *Phys. Rev. Lett.*, 2016, **117**, 010503.
66 S. Paesani, A. A. Gentile, R. Santagati, J. Wang, N. Wiebe, D. P. Tew, J. L. O'Brien and M. G. Thompson, Experimental Bayesian quantum phase estimation on a silicon photonic chip, *Phys. Rev. Lett.*, 2017, **118**, 100503.
67 T. E. O'Brien, B. Tarasinski and B. M. Terhal, Quantum phase estimation of multiple eigenvalues for small-scale (noisy) experiments, *New J. Phys.*, 2019, **21**, 023022.
68 V. G. Gebhart, A. Smerzi and L. Pezzè, Bayesian Quantum multiphase estimation algorithm, *Phys. Rev. A*, 2021, **16**, 014035.
69 S. Kimmel, G. H. Low and T. J. Yoder, Robust calibration of a universal single-qubit gate set via robust phase estimation, *Phys. Rev. A*, 2015, **92**, 062315.
70 E. van den Berg, Iterative quantum phase estimation with optimized sample complexity, 2020 IEEE International Conference on Quantum Computing and Engineering (QCE), Denver, CO, USA, 2020, pp. 1-10
71 A. D. Becke, Density-functional thermochemistry. III. The role of exact exchange, *J. Chem. Phys.*, 1993, **98**, 5648–5652.
72 C. Lee, W. Yang and R. G. Parr, Development of the Colle-Salvetti correlation-energy formula into a functional of the electron density, *Phys. Rev. B*, 1988, **37**, 785–789.
73 S. H. Vosko, L. Wilk and M. Nusair, Accurate spin-dependent electron liquid correlation energies for local spin density calculations: a critical analysis, *Can. J. Phys.*, 1980, **58,** 1200–1211.
74 M. J. Frisch, G. W. Trucks, H. B. Schlegel, G. E. Scuseria, M. A. Robb, J. R. Cheeseman, G. Scalmani, V. Barone, G. A. Petersson, H. Nakatsuji, X. Li, M. Caricato, A. V. Marenich, J. Bloino, B. G. Janesko, R. Gomperts, B. Mennucci, H. P. Hratchian, J. V. Ortiz, A. F. Izmaylov, J. L. Sonnenberg, D. Williams-Young, F. Ding, F. Lipparini, F. Egidi, J. Goings, B. Peng, A. Petrone, T. Henderson, D. Ranasinghe, V. G. Zakrzewski, J. Gao, N. Rega, G. Zheng, W. Liang, M. Hada, M. Ehara, K. Toyota, R. Fukuda, J. Hasegawa, M. Ishida, T. Nakajima, Y. Honda, O. Kitao, H. Nakai, T. Vreven, K. Throssell, J. A. Montgomery, Jr., J. E. Peralta, F. Ogliaro, M. J. Bearpark, J. J. Heyd, E. N. Brothers, K. N. Kudin, V. N. Staroverov, T. A. Keith, R. Kobayashi, J. Normand, K. Raghavachari, A. P. Rendell, J. C. Burant, S. S. Iyengar, J. Tomasi, M. Cossi, J. M. Millam, M. Klene, C. Adamo, R. Cammi, J. W. Ochterski, R. L. Martin, K. Morokuma, O. Farkas, J. B. Foresman and D. J. Fox, Gaussian 16 (Revision B.01), Gaussian Inc., Wallingford CT, 2016
75 T. Hashimoto, H. Nakano and K. Hirao, Theoretical study of the valence π–π* excited states of polyacenes: benzene and naphthalene, *J. Chem. Phys.*, 1996, **104,** 6244–6258.
76 Y. J. Liu, P. Persson and S. Lunell, Multireference calculations of the phosphorescence and photodissociation of chlorobenzene, *J. Chem. Phys.*, 2004, **121,** 11000–11006.
77 J. Soto and M. Algarra, Electronic structure of nitrobenzene: a benchmark example of the accuracy of the multi-state CASPT2 theory, *J. Phys. Chem. A*, 2021, **125**, 9431–9437.
78 H. F. Trotter, On the product of semi-groups of operators, *Proc. Am. Math. Soc.,* 1959, **10**, 545–551.
79 M. Suzuki, Generalized Trotter's formula and systematic approximants of exponential operators and inner derivations with applications to many-body problems, *Commun. Math. Phys.*, 1976, **51**, 183–190.

80  Q. Sun, X. Zhang, S. Banerjee, P. Bao, M. Barbry, N. S. Blunt, N. A. Bogdanov, G. H. Booth, J. Chen, Z.-H. Cui, J. J. Eriksen, Y. Gao, S. Guo, J. Hermann, M. R. Hermes, K. Koh, P. Koval, S. Lehtola, Z. Li, J. Liu, N. Mardirossian, J. D. McClain, M. Motta, B. Mussard, H. Q. Pham, A. Pulkin, W. Purwanto, P. J. Robinson, E. Ronca, E. R. Sayfutyarova, M. Scheurer, H. F. Schurkus, J. E. T. Smith, C. Sun, S.-N. Sun, S. Upadhyay, L. K. Wagner, X. Wang, A. White, J. D. Whitfield, M. J. Williamson, S. Wouters, J. Yang, J. M. Yu, T. Zhu, T. C. Berkelbach, S. Sharma, A. Y. Sokolov and G. K.-L. Chan, Recent developments in the PySCF program package, *J. Chem. Phys.*, 2020, **153**, 024109.
81  Cirq Developers. Cirq (v1.0.0), Google Quantum AI, Santa Barbara, CA, 2023. DOI: 10.5281/zenodo.4062499
82  J. R McClean, N. C. Rubin, K. J. Sung, I. D. Kivlichan, X. Bonet-Monroig, Y. Cao, C. Dai, E. S. Fried, C. Gidney, B. Gimby, P. Gokhale, T. Häner, T. Hardikar, V. Havlíček, O. Higgott, C. Huang, J. Izaac, Z. Jiang, X. Liu, S. McArdle, M. Neeley, T. O'Brien, B. O'Gorman, I. Ozfidan, M. D. Radin, J. Romero, N. P. D. Sawaya, B. Senjean, K. Setia, S. Sim, D. S. Steiger, M. Steudtner, Q. Sun, W. Sun, D. Wang, F. Zhang and R. Babbush, OpenFermion: the electronic structure package for quantum computers, *Quantum Sci. Technol.*, 2020, **5**, 034014.
83  Quantum AI team and collaborators. qsim. 2020, Google Quantum AI, Santa Barbara, CA, 2023. DOI: 10.5281/zenodo.4023103
84  Nvidia Corporation Home Page, https://blogs.nvidia.com/blog/2018/09/12/fujifilm-adopts-nvidia-dgx-2/ (accessed 2023-10-24)
85  K. Mitarai, K. Toyoizumi and W. Mizukami, Perturbation theory with quantum signal processing, *Quantum*, 2023, **7**, 1000.
86  C. L. Cortes, M. Loipersberger, R. M. Parrish, S. Morley-Short, W. Pol, S. Sim, M. Steudtner, C. S. Tautermann, M. Degroote, N. Moll, R. Santagati and M. Streif, Fault-tolerant quantum algorithm for symmetry-adapted perturbation theory, *PRX Quantum*, 2024, **5**, 010336
87  S. Bravyi, J. M. Gambetta, A. Mezzacapo and K. Temme, Tapering off qubits to simulate fermionic Hamiltonians, *arXiv*, 2017, preprint, arXiv:1701.08213, https://arxiv.org/abs/1701.08213
88  K. Setia, R. Chen, J. E. Rice, A. Mezzacapo, M. Pistoia and J. D. Whitfield, Reducing qubit requirements for quantum simulations using molecular point group symmetries, *J. Chem. Theory Comput.*, 2020, **16,** 6091–6097.
89  M. B. Hastings, D. Wecker, B. Bauer and M. Troyer, Improving quantum algorithms for quantum chemistry, *Quantum Info. Comput.*, 2015, **15**, 1–21.

# Workflow for practical quantum chemical calculations with quantum phase estimation algorithm: electronic ground and π-π* excited states of benzene and its derivatives

Yusuke Ino* †, Misaki Yonekawa †, Hideto Yuzawa †, Yuichiro Minato ‡, and Kenji Sugisaki* ¶

† Fujifilm Corporation, 577 Ushijima, Kaisei-cho, Ashigarakami-gun, Kanagawa 258-8577, Japan,

‡ blueqat Inc., 2-24-12-39F, Shibuya, Shibuya-ku, Tokyo 150-6139, Japan,

¶ Graduate School of Science and Technology, Keio University, 7-1 Shinkawasaki, Saiwai-ku, Kawasaki, Kanagawa 212-0032, Japan; Quantum Computing Center, Keio University, 3-14-1 Hiyoshi, Kohoku-ku Yokohama, Kanagawa 223-8522, Japan; Centre for Quantum Engineering, Research and Education, TCG Centres for Research and Education in Science and Technology, Sector V, Salt Lake, Kolkata 700091, India

1. Cartesian coordinates of LiH, benzene, chlorobenzene, nitrobenzene

Table S.1.1 Cartesian coordinates of LiH molecule optimized at the B3LYP/6-31+G(d) level of theory.

| Atom | x | y | z |
|---|---|---|---|
| H | 0 | 0 | -1.1443 |
| Li | 0 | 0 | 0.3814 |

Table S.1.2 Cartesian coordinates of benzene optimized at the B3LYP/6-311G(d) level of theory.

| Atom | x | y | z |
|---|---|---|---|
| C | 0.0000 | -1.3966 | 0.0000 |
| C | -1.2095 | -0.6983 | 0.0000 |
| C | 1.2095 | -0.6983 | 0.0000 |
| C | 1.2095 | 0.6983 | 0.0000 |
| C | -1.2095 | 0.6983 | 0.0000 |
| C | 0.0000 | 1.3966 | 0.0000 |
| H | 0.0000 | -2.4837 | 0.0000 |
| H | 2.1509 | -1.2418 | 0.0000 |
| H | 2.1509 | 1.2418 | 0.0000 |
| H | -2.1509 | -1.2418 | 0.0000 |
| H | -2.1509 | 1.2418 | 0.0000 |
| H | 0.0000 | 2.4837 | 0.0000 |

Table S.1.3 Cartesian coordinates of chlorobenzene optimized at the B3LYP/6-311G(d) level of theory.

| Atom | x | y | z |
|---|---|---|---|
| C | 0.0000 | -1.2163 | -0.1783 |
| C | 0.0000 | 0.0000 | 0.5034 |
| C | 0.0000 | -1.2076 | -1.5743 |
| C | 0.0000 | 0.0000 | -2.2745 |
| C | 0.0000 | 1.2163 | -0.1783 |
| C | 0.0000 | 1.2076 | -1.5743 |
| H | 0.0000 | -2.1514 | -2.1128 |
| H | 0.0000 | 0.0000 | -3.3607 |
| H | 0.0000 | -2.1495 | 0.3749 |
| H | 0.0000 | 2.1495 | 0.3749 |
| H | 0.0000 | 2.1514 | -2.1128 |
| Cl | 0.0000 | 0.0000 | 2.2644 |

Table S.1.4 Cartesian coordinates of nitrobenzene optimized at the B3LYP/6-311G(d) level of theory.

| Atom | x | y | z |
|---|---|---|---|
| C | 0.0000 | 0.0000 | -0.2450 |
| C | 0.0000 | -1.2207 | 0.4278 |
| C | 0.0000 | 1.2207 | 0.4278 |
| C | 0.0000 | 1.2123 | 1.8209 |
| C | 0.0000 | -1.2123 | 1.8209 |
| C | 0.0000 | 0.0000 | 2.5164 |
| H | 0.0000 | 2.1429 | -0.1400 |
| H | 0.0000 | 2.1532 | 2.3631 |
| H | 0.0000 | -2.1429 | -0.1400 |
| H | 0.0000 | -2.1532 | 2.3631 |
| H | 0.0000 | 0.0000 | 3.6028 |
| N | 0.0000 | 0.0000 | -1.7179 |
| O | 0.0000 | -1.0898 | -2.2898 |
| O | 0.0000 | 1.0898 | -2.2898 |

2. Active orbitals used for the CAS-CI calculations of Benzene, Chlorobenzene, Nitrobenzene

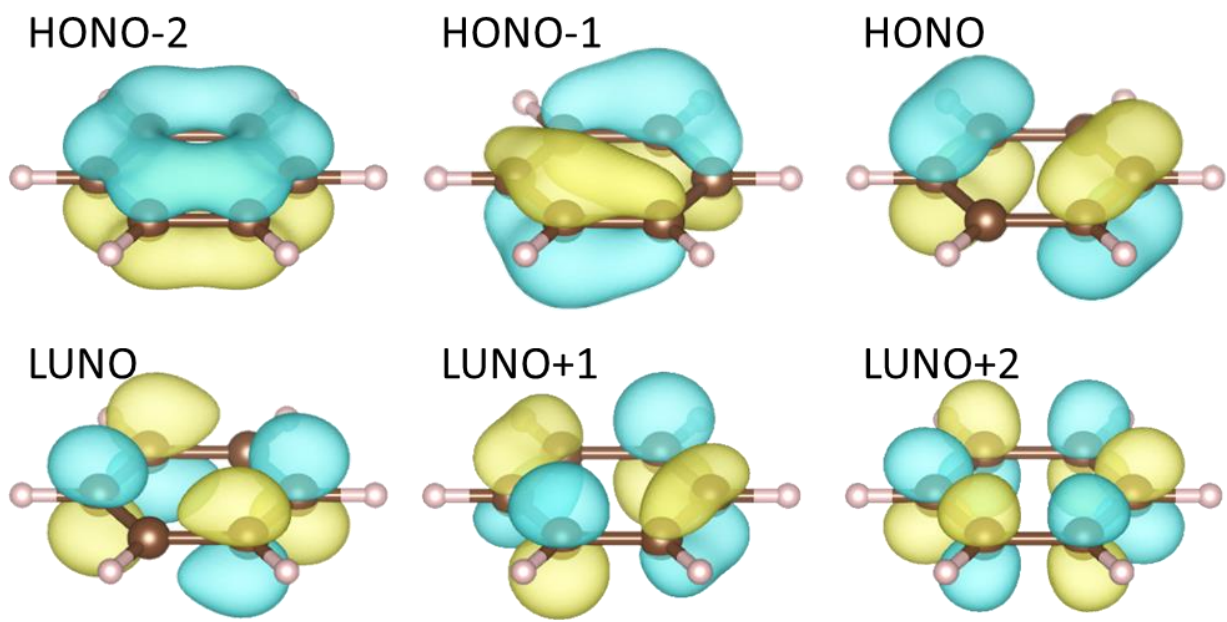

Figure S.2.1. Active orbitals used for the CAS-CI calculations of benzene. Label at the upper left of each orbital refers to the index of the pseudo-natural orbitals constructed from the MP2 calculation.

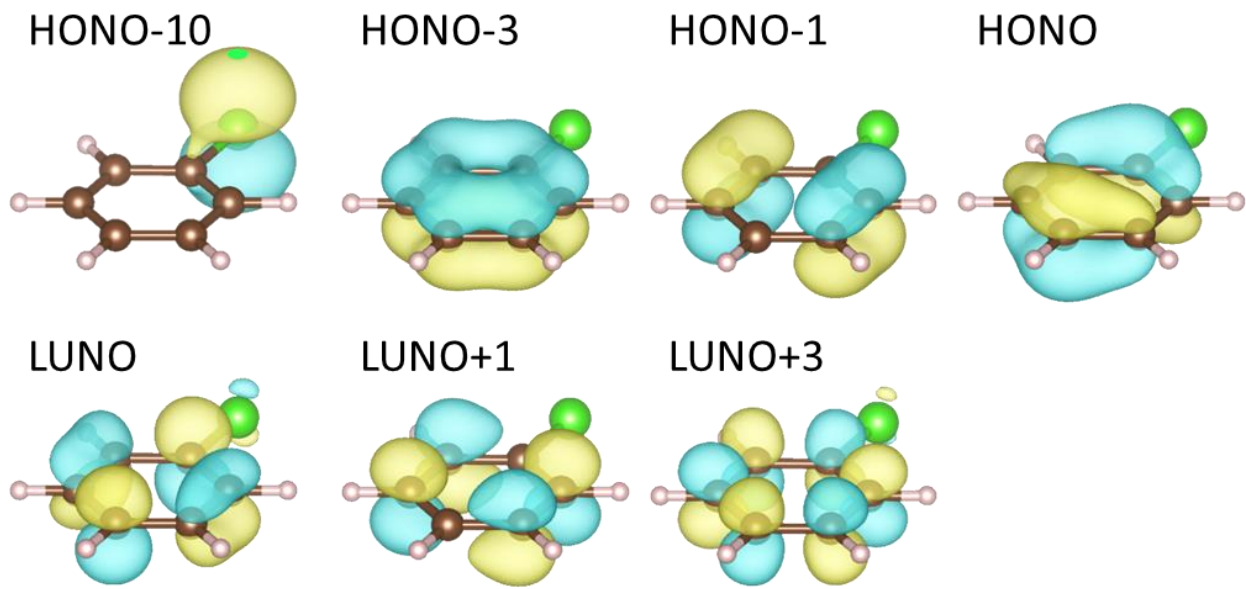

Figure S.2.2. Active orbitals used for the CAS-CI calculations of chlorobenzene. Label at the upper left of each orbital refers to the index of the pseudo-natural orbitals constructed from the MP2 calculation.

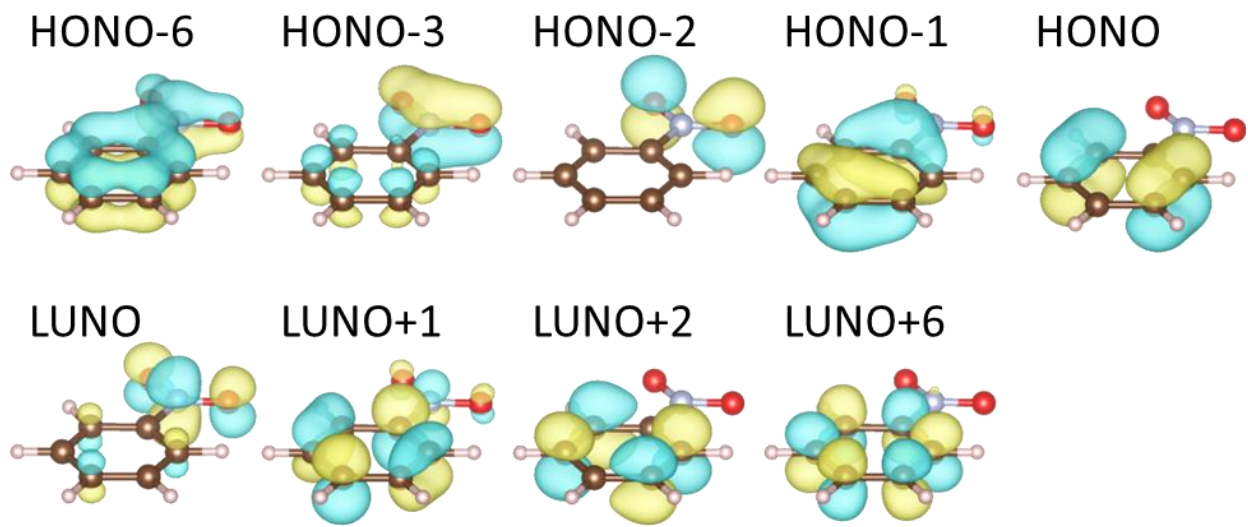

Figure S.2.3. Active orbitals used for the CAS-CI calculations of nitrobenzene. Label at the upper left of each orbital refers to the index of the pseudo-natural orbitals constructed from the MP2 calculation.

3. Details of quantum phase estimation

Table S.3.1 The detail of Hamiltonian and quantum phase estimation for LiH.

| Number of qubits | 13 | 15 | 17 | 19 | 21 |
|---|---|---|---|---|---|
| Number of electrons | 4 | 4 | 4 | 4 | 4 |
| Number of atoms | 2 | 2 | 2 | 2 | 2 |
| Number of active spin-orbitals | 12 | 14 | 16 | 18 | 20 |
| Number of one-electron integrals | 36 | 54 | 76 | 78 | 80 |
| Number of two-electron integrals | 1824 | 3732 | 6944 | 9012 | 11168 |
| Number of quantum gates (for determining the whole 4 digits) | 3888 | 9348 | 19684 | 26304 | 33884 |

Table S.3.2 The detail of Hamiltonian and quantum phase estimation for benzene, chlorobenzene, nitrobenzene

| | Benzene | Chlorobenzene | Nitrobenzene |
|---|---|---|---|
| Number of electrons | 42 | 58 | 64 |
| Number of atoms | 12 | 12 | 14 |
| Number of active spin-orbitals | 12 | 14 | 18 |
| Number of one-electron integrals | 16 | 58 | 90 |
| Number of two-electron integrals | 1184 | 4964 | 23284 |
| Number of qubits | 13 | 15 | 19 |
| Number of quantum gates (for determining the whole 10 digits) | 2420 | 12624 | 42156 |

4. Quantum Circuits used for the preparation of the excited state approximate wave function.

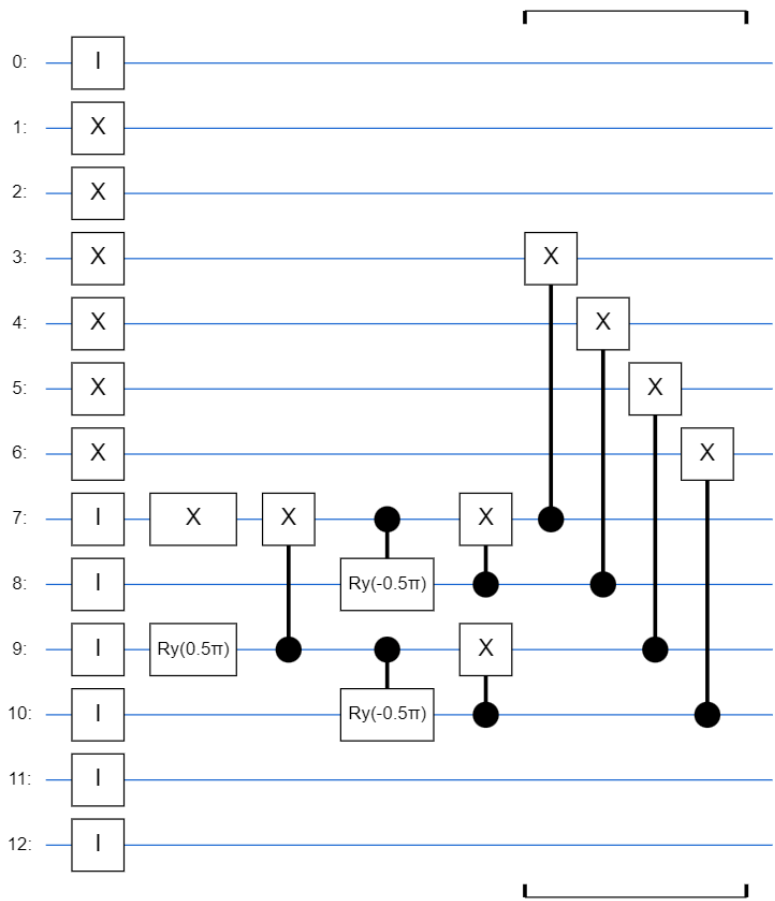

Figure S.4.1 Quantum circuit used for the preparation of the excited state approximate wave function of benzene.

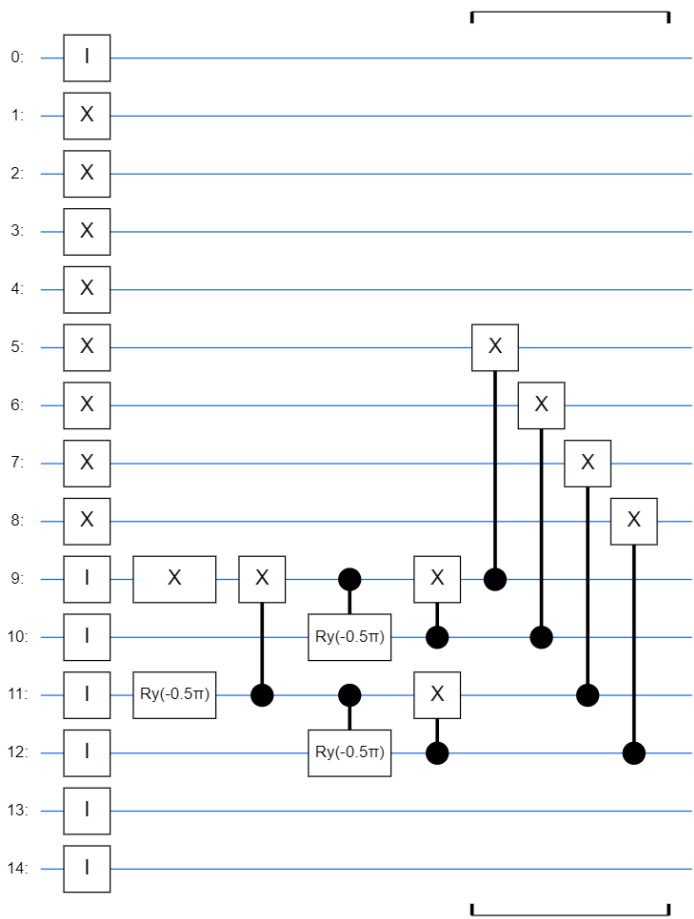

Figure S.4.2 Quantum circuit used for the preparation of the excited state approximate wave function of chlorobenzene.

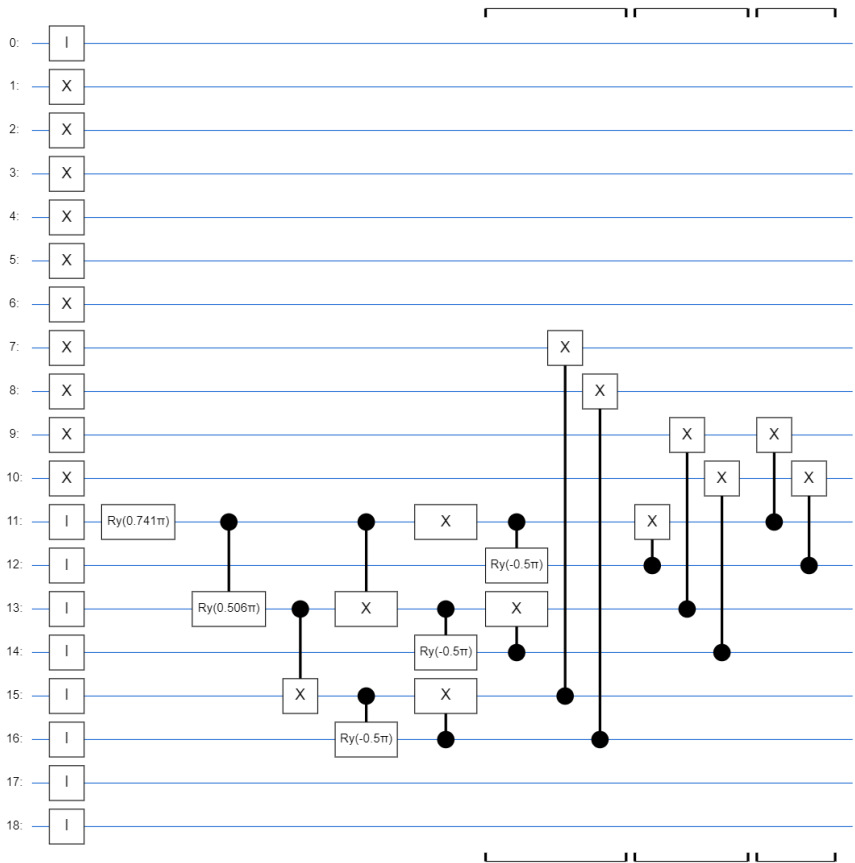

Figure S.4.3 Quantum circuit used for the preparation of the excited state approximate wave function of benzene.